\documentclass[
    reprint,
    aps,
    prc,
    floatfix,
    superscriptaddress,
    longbibliography
]{revtex4-2}

\usepackage[T1]{fontenc}
\usepackage{physics}
\usepackage{graphicx}
\usepackage{dcolumn}
\usepackage{bm}
\usepackage{booktabs}
\usepackage{makecell}
\usepackage{siunitx}
\usepackage[hidelinks]{hyperref}
\usepackage{cleveref}
\usepackage{orcidlink} 
\begin{document}

\title{Non-perturbative charm production from color ropes in relativistic heavy ion collisions}
\author{Carl B. Rosenkvist\,\orcidlink{0009-0001-8384-2806}\,}
\email{rosenkvist@itp.uni-frankfurt.de}
\affiliation{Frankfurt Institute for Advanced Studies, Ruth-Moufang-Straße 1, 60438 Frankfurt am Main, Germany}
\affiliation{Institut für Theoretische Physik, Goethe-Universität Frankfurt, Max-von-Laue-Straße 1, 60438 Frankfurt am Main, Germany}
\affiliation{GSI Helmholtzzentrum für Schwerionenforschung GmbH, Planckstraße 1, 64291 Darmstadt, Germany}

\author{Marcus Bleicher\,\orcidlink{0000-0003-2926-642X}\,}
\affiliation{Institut für Theoretische Physik, Goethe-Universität Frankfurt, Max-von-Laue-Straße 1, 60438 Frankfurt am Main, Germany}
\affiliation{Helmholtz Research Academy Hesse for FAIR, GSI Helmholtz Center, Campus Frankfurt, Max-von-Laue-Straße 12, 60438 Frankfurt am Main, Germany}
\date{\today}

\begin{abstract}
We present results on charm production in nucleus--nucleus reactions in the energy range $\sqrt{s_{NN}}=5$--$20$~GeV, obtained with the SMASH transport approach extended to include color rope formation.
We show that string fusion may provide a substantial additional non-perturbative source of charm production. This may help to explain the apparent charm enhancement (above the pQCD expectations) at low transverse momenta. Our predicted neutral $D$-meson yields are consistent with current experimental constraints and may provide valuable insights for upcoming data by the NA61/SHINE collaboration.
\end{abstract}

\maketitle

\section{Introduction}
Relativistic heavy-ion collisions provide access to strongly interacting matter at high energy and baryon densities and allows one to explore the properties of quantum chromodynamics (QCD) under extreme conditions~\cite{Sorensen:2023zkk}. Heavy-flavor particles are particularly valuable probes of the early collision dynamics as they are typically produced under controlled (pQCD) conditions in the initial stage. In the conventional description, charm quarks are produced predominantly in initial hard partonic scatterings, while their subsequent propagation and hadronization reflect their interaction with the surrounding matter~\cite{Andronic:2015wma}. Typically, at high energies initial charm production is dominated by gluon fusion. However, leading-order calculations generally need to be corrected by a phenomenological K-factor, whose magnitude depends on the collision energy, the parton distributions, and further parameters~\cite{Vogt:2002eu}. Going beyond leading order (see e.g. the FONLL calculations in~\cite{Cacciari:2012ny}) reduces the discrepancies with the data substantially. Nevertheless, recent ALICE measurements of charm production~\cite{ALICE:2023sgl} lie systematically at the upper edge of the FONLL uncertainty band, and at low transverse momenta exceed the central FONLL prediction by up to a factor of about two. Together with the large theoretical uncertainties in this region, this indicates that the mechanisms producing the initial charm quarks, especially in the low $p_T$ region, are not yet fully understood, although they are relevant for the interpretation of open-charm and charmonium observables. Let us also note that perturbative QCD calculations become less reliable at lower collision energies, e.g. in the CERN-SPS energy range or the low RHIC-BES regime. In this paper, we propose an additional source of charm production due to the formation of strong color fields, which would modify the production baseline relative to the pQCD expectation and yield additional charm quarks at low transverse momenta.

In string-based descriptions of hadronic and nuclear reactions, particle production proceeds through the excitation and fragmentation of color flux tubes (called strings). Usually, these strings are treated as independent objects. While this is a good assumption in low energetic collisions or low multiplicity events at high energies, it becomes questionable when many strings occupy the same space--time region. It is well known that in such a setting, the individual color fields can overlap and combine into configurations carrying higher color representations, commonly referred to as color ropes~\cite{Biro:1984cf} (also called string fusion). The resulting increase of the effective string tension changes the flavor composition of the particles produced during fragmentation. This mechanism is well known and was investigated in the past for hadron-nucleus reactions~\cite{Armesto:1994er}, nucleus-nucleus reactions at SPS~\cite{Armesto:1994yg,Sorge:1995dp} and RHIC energies~\cite{Bleicher:2000us}. The typical set-up in these exploratory studies was to increase the string tension from $1\,\mathrm{GeV/fm}$ to $3\,\mathrm{GeV/fm}$ which leads to an enhanced strange-quark and diquark production and generated a pronounced hierarchy of hyperon enhancement with increasing strangeness content. These results illustrated that the enhanced production of massive constituents can originate from collective color fields without requiring chemical equilibration in a thermal plasma.

The reason for this enhancement follows directly from the Schwinger mechanism for pair creation in an electric field~\cite{Schwinger:1951nm}, which has also been applied to quark–antiquark production in chromo-electric flux tubes~\cite{Casher:1978wy}. In its simplest application to string fragmentation~\cite{Andersson:1983ia}, the relative probability to produce a heavy quark flavor $Q$, compared with a light flavor $q$, is commonly written as

\begin{equation}
\gamma_Q = \frac{P(Q\overline Q)}{P(q\overline q)}=
\exp\left[-\frac{\pi(m_Q^2-m_q^2)}{\kappa_{\mathrm{eff}}}\right],
\label{eq:schwinger}
\end{equation}
where $m_Q$ and $m_q$ denote the heavy and light quark masses entering the fragmentation model, and $\kappa_{\mathrm{eff}}$ is the effective tension governing pair creation. For the tension of an isolated vacuum string, charm production through tunneling is very strongly suppressed. An increase in $\kappa_{\mathrm{eff}}$ exponentially reduces this suppression, making charm production, in addition to the well-studied strangeness and antibaryon sectors, a particularly sensitive probe of the field strengths associated with overlapping strings.

As we will see below, a tenfold increase in the string tension will increase the relative probability for charm production in the string fragmentation process by approximately ten orders of magnitude from ${\mathcal{P}}(c\bar c)(\kappa=1 {\rm GeV/fm}) \sim 10^{-11}$ to ${\mathcal{P}}(c\bar c) (\kappa=10 {\rm GeV/fm}) \sim 10^{-1}$.

The possibility of enhanced charm production due to the increased string tension associated with rope formation was already discussed in Ref.~\cite{Levai:2000ne} within the ALCOR coalescence model. However, the enhancement was only estimated via the Schwinger formula, while the total charm yield was fixed by fitting the measured $J/\psi$ to $h^-$ ratio rather than calculated dynamically.

Later, enhanced charm production from strong longitudinal color fields was explored within the HIJING/$B\overline B$ model~\cite{ToporPop:2009dpl}, using more realistic string configurations. These pioneering calculations demonstrated that an increased effective string tension can indeed generate a substantial non-perturbative contribution to open charm production. 

Since the earlier QGSM/RQMD/UrQMD studies, string fusion and rope formation have been developed in several complementary directions. The QGSM string fusion model incorporated the formation and fragmentation of higher-color strings into Monte Carlo simulations of nuclear collisions~\cite{Amelin:2001sk}. String-percolation approaches extended the treatment to clusters of overlapping color sources, connecting the non-Abelian addition of color charges to changes in multiplicities and transverse-momentum distributions~\cite{Braun:2015eoa}. In a different approach, the Generalized Effective String Rope Model described the initial stages of relativistic heavy-ion collisions in terms of strong longitudinal chromoelectric fields, with effective string tensions substantially exceeding those of ordinary hadronic strings~\cite{Ramirez:2022obv}. 

Rope hadronization has also been incorporated into PYTHIA/Angantyr~\cite{bierlich2022comprehensiveguidephysicsusage,Angantyr} to investigate collective phenomena across collision systems, from proton–proton to heavy-ion collisions, without requiring the formation of a thermalized medium~\cite{Bierlich:2014xba,Bierlich:2017sxk,Bierlich:2022ned,Bierlich:2022oja}. More recently, the SMASH transport framework has been extended to include rope hadronization (SMASH+Ropes), providing a complementary approach to studying strangeness production in non-equilibrium systems~\cite{Rosenkvist:2026wyp}.

Recently, ALICE data further suggests that the relevance of interacting strings extends beyond nucleus--nucleus collisions. In fact, ALICE observed increasing strange- and multistrange-hadron yields relative to pions with charged-particle multiplicity even in proton--proton collisions~\cite{ALICE:2016fzo}. As suggested in \cite{Bierlich:2022ned}, this enhancement might also be interpreted in terms of a higher effective string tension or rope formation in high-multiplicity pp or pA reactions. 

In the present work, we investigate charm production in SMASH+Ropes by extending the strong-field production argument from strangeness to charm and examine whether the color fields generated by overlapping strings can provide an additional relevant source of $c\overline c$ pairs. The central quantitative question is whether the enhanced tunneling probability in case of ropes can be realized with the available energy and if the space-time dynamics allows to create a sufficient abundance of suitable rope configurations to affect the total charm yield. This investigation provides a test of the heavy-flavor consequences of collective string dynamics and of their relevance to the interpretation of charm production in dense collision environments.

\section{Models}
Here we summarize the models used in this work and briefly describe the relevant physics for charm production in these approaches. We use the rope approach SMASH+Ropes~\cite{Rosenkvist:2026wyp}
which is implemented into the transport model SMASH~\cite{SMASH:2016zqf}. SMASH employs Pythia~\cite{bierlich2022comprehensiveguidephysicsusage} to model string fragmentation and hard scatterings. The specific model versions used in this work can be found in the appendix section.

\subsection{SMASH}
Simulating Many Accelerated Strongly-interacting Hadrons (SMASH)~\cite{SMASH:2016zqf} is a relativistic hadronic transport model that includes all well-established hadrons with masses up to approximately $2~\mathrm{GeV}$ as active degrees of freedom. The dynamical evolution is governed by the relativistic Boltzmann equation,
\begin{equation}
p^\mu \partial_\mu f_i(x,p) + m_i F^\alpha \partial_\alpha^p f_i(x,p) = C_\text{coll}^i,
\end{equation}
where $f_i(x,p)$ is the single-particle distribution function for species $i$, $C_\text{coll}^i$ is the collision term describing elastic and inelastic interactions, as well as resonance formation and decays. $F^\alpha$ represents mean-field potentials, which are neglected in this work. The collision criterion uses the geometrical interpretation of the cross section, with an interaction radius determined by the total cross section, $d_\text{int}=\sqrt{\sigma_\text{tot}/\pi}$.

At lower collision energies, inelastic interactions and hadron production are modeled through resonance formation and decay. With increasing collision energy, these processes are gradually replaced by string excitations, with a transition between $\sqrt{s}=\SI{3.5}{GeV}$ and $\SI{4.5}{GeV}$. Soft string excitations are divided into diffractive and non-diffractive processes. Diffractive interactions involve the excitation of one or both incoming hadrons into strings, while non-diffractive interactions involve color exchange through valence quark exchange.

At higher collision energies, SMASH employs Pythia's semi-perturbative multiparton interaction framework to describe hard string excitations, following the implementation of Ref.~\cite{Mohs:2019iee}. Assuming a Poisson distribution for the number of hard interactions, the hard non-diffractive cross section is given by
\begin{equation}
\sigma_{\mathrm{ND,hard}}
=
\sigma_{\mathrm{ND}}
\left(
1-\exp\left[-\frac{\sigma_{\mathrm{hard}}}{\sigma_{\mathrm{ND}}}\right]
\right),
\end{equation}
where $\sigma_{\mathrm{hard}}$ is the perturbative hard-scattering cross section and $\sigma_{\mathrm{ND}}$ is the total non-diffractive cross section.

\begin{figure}[t]
    \centering
    \includegraphics[width=\columnwidth]{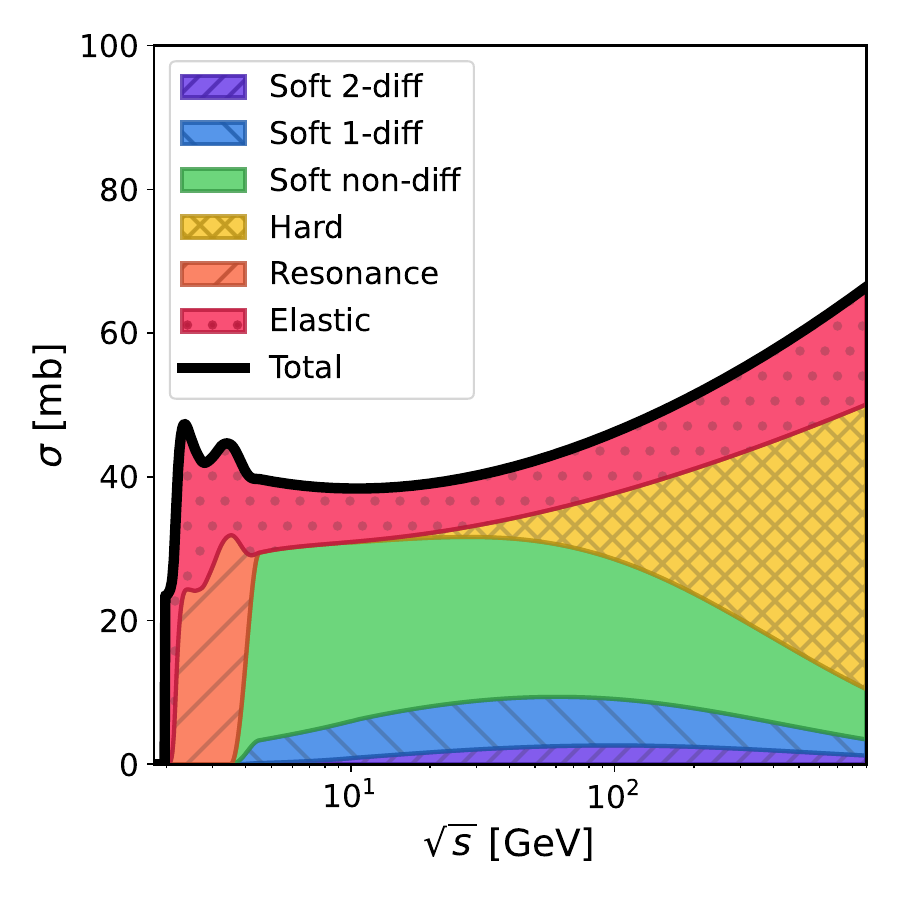}
    \caption{Center-of-mass energy dependence of the proton--proton cross sections calculated with SMASH. The stacked shaded areas show the contributions from elastic scattering, resonance excitation, and soft and hard string excitation. Soft string excitations are separated into single- and double-diffractive and non-diffractive processes. The solid black line represents the total cross section.}
    \label{fig:tot_xs}
\end{figure}

The contributions of the different interaction processes to the total proton--proton cross section are shown in Fig.~\ref{fig:tot_xs}. At lower energies, resonance excitation dominates the inelastic contributions, while soft string excitations become increasingly important as the collision energy increases. The hard string contribution remains negligible at SPS energies, where soft string excitations dominate string production.

All strings produced in SMASH are fragmented using Pythia, independently of their production mechanism.

\subsection{Pythia}\label{sec:pythia}
Pythia is a general-purpose event generator for hadronic collisions, combining perturbative and phenomenological descriptions across different momentum scales. Since all string configurations in SMASH are fragmented by Pythia, let us briefly summarize the aspects relevant to this work, following Ref.~\cite{bierlich2022comprehensiveguidephysicsusage}.

An important component of Pythia is the multiparton interaction (MPI) framework, which describes multiple semi-perturbative parton--parton scatterings within a single event~\cite{Sjostrand:1987su}. Hadronization is modeled using the Lund string model, in which the color field between partons is represented by a relativistic string with a constant energy density. The string fragments through the tunneling production of quark--antiquark pairs.

The probability $\mathcal{P}$ of producing a quark--antiquark pair with quark mass $m_q$ and a transverse momentum $p_\perp$ during string breaking is given by
\begin{equation}\label{eq:tunneling_string}
\frac{1}{\kappa} \frac{d\mathcal{P}}{dp_\perp}
\propto
\exp\left(-\frac{\pi m_\perp^2}{\kappa}\right),
\end{equation}
where $\kappa$ is the string tension and $m_\perp=\sqrt{m_q^2+p^2_\perp}$. Consequently, the relative probability of producing a quark--antiquark pair of flavor $q$ compared to a light quark--antiquark pair is suppressed by the quark mass:
\begin{equation}\label{eq:mass_supp_string}
\frac{\mathcal{P}_q}{\mathcal{P}_{u,d}}
=
\exp\left(-\pi\frac{m_q^2-m_u^2}{\kappa}\right).
\end{equation}

In Pythia, the suppression of strange quark--antiquark pair production is tuned to experimental data, while charm production is calculated from the mass of the charm quark and is strongly suppressed due to the high charm quark mass. The default flavor suppression settings are
\begin{equation}
\mathcal{P}(u\bar u):\mathcal{P}(d\bar d):
\mathcal{P}(s\bar s):\mathcal{P}(c\bar c)
=1:1:0.08:10^{-11}.
\end{equation}

Due to this strong suppression, charm production through standard string fragmentation is usually neglected, although charm quarks produced in hard scatterings may be present at the string endpoints.

\subsection{SMASH+Ropes}
SMASH+Ropes was developed to investigate non-equilibrium strangeness enhancement at intermediate collision energies and to compare its predictions with those of thermal models, including viscous corrections~\cite{Rosenkvist:2026wyp}. The model is based on the idea that delaying string fragmentation provides access to the space--time evolution of strings and therefore allows the local overlap of strings to be estimated dynamically.

In conventional transport models, baryon stopping processes are difficult to describe rigorously because the system evolves in time and newly produced hadrons are often treated as unformed particles with reduced interaction cross sections. In SMASH+Ropes, however, the stopping mechanism is carried by the strings themselves. This is achieved by allowing strings to interact with surrounding hadrons through diffractive excitations before they fragment.

The probability of rope formation in a cell $c$ of the SMASH interaction grid is defined as
\begin{equation}
P_{\mathrm{rope}}(c)
=
\min\left(1,\frac{V_{\mathrm{tot,string}}(c)}{V_c}\right),
\end{equation}
where $V_{\mathrm{tot,string}}(c)=\sum_{i\in c}V_i$ is the total string volume contained within the cell and $V_c$ denotes the cell volume. Each string is assigned a constant effective volume,
\begin{equation}
V_i \equiv V_{\mathrm{string}}.
\end{equation}
The rope multiplicity is determined using the same SU(3) random-walk prescription employed in Pythia/Angantyr, following Ref.~\cite{Bierlich:2014xba}. The resulting color multiplet transition,
\begin{equation}
\{p_{\mathrm{before}}, q_{\mathrm{before}}\}
\rightarrow
\{p_{\mathrm{after}}, q_{\mathrm{after}}\},
\end{equation}
is then used to calculate the effective rope tension. Here, $p$ and $q$ are the Dynkin labels of the irreducible SU(3) color representation; they are binomially distributed according to the number of strings. The string tension is assumed to scale with the quadratic Casimir operator~\cite{Biro:1984cf},
\begin{equation}
\frac{\kappa^{\{p,q\}}}{\kappa^{\{1,0\}}}
=
\frac{C_2(p,q)}{C_2(1,0)}
=
\frac{1}{4}
\left(
p^2 + q^2 + pq + 3(p+q)
\right).
\end{equation}

\begin{figure}[t]
\centering
\includegraphics[width=\columnwidth]{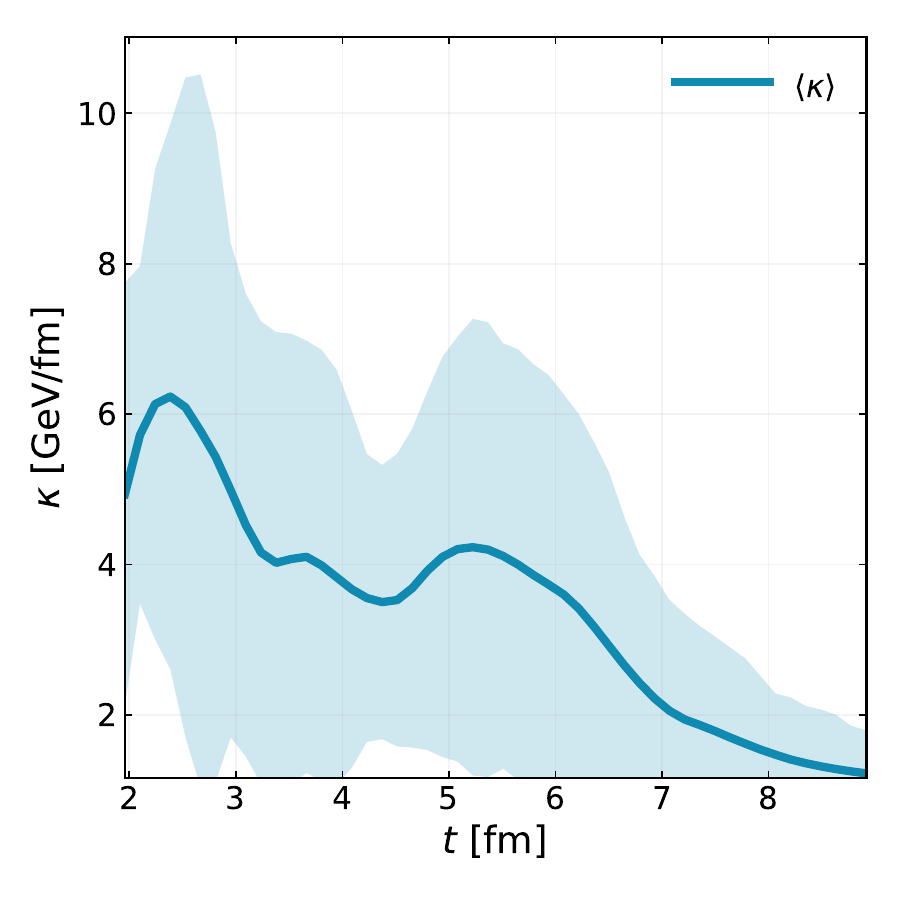}
\caption{
Time evolution of the mean effective string tension $\kappa_{\mathrm{eff}}$ at fragmentation in a central slice of central Pb--Pb collisions at $\sqrt{s_{NN}}=17.3$~GeV, obtained with $V_\text{string}=\SI{3.15}{fm^3}$. The shaded band shows the standard deviation of $\kappa_{\mathrm{eff}}$ over all fragmenting strings.
}
\label{fig:k_eff_time_parm}
\end{figure}

As an illustration of the resulting rope dynamics, Fig.~\ref{fig:k_eff_time_parm} shows the time evolution of the average effective string tension in central Pb--Pb collisions. The tension reaches its maximum during the densest stage of the collision, where string overlap is largest, before decreasing as the system expands and the strings fragment. One observes that the effective string tension can reach substantially higher values (even up to $10$~GeV/fm) than the tension of about $1$~GeV/fm of a single string.

\section{Constraints on Charm Enhancement}\label{sec:charm_constraints}
\begin{figure}[t]
\centering
\includegraphics[width=\columnwidth]{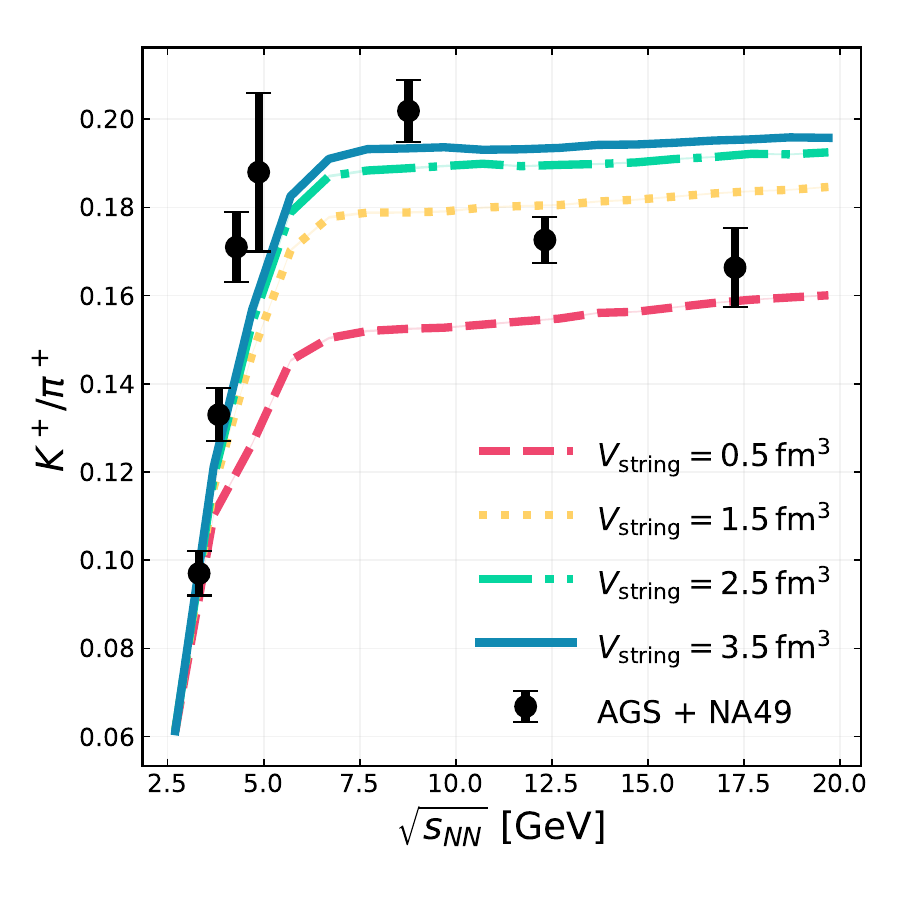}
\caption{Collision-energy dependence of the ratio of mean
$K^+$ and $\pi^+$ multiplicities in full phase space
for central Pb–Pb collisions. The different curves
correspond to different string volumes.
Experimental data from AGS and NA49 are taken from
Ref.~\cite{NA49:2002pzu}.}
\label{fig:k_pi_plus-volume-scan}
\end{figure}

\begin{figure}[t]
\centering
\includegraphics[width=\columnwidth]{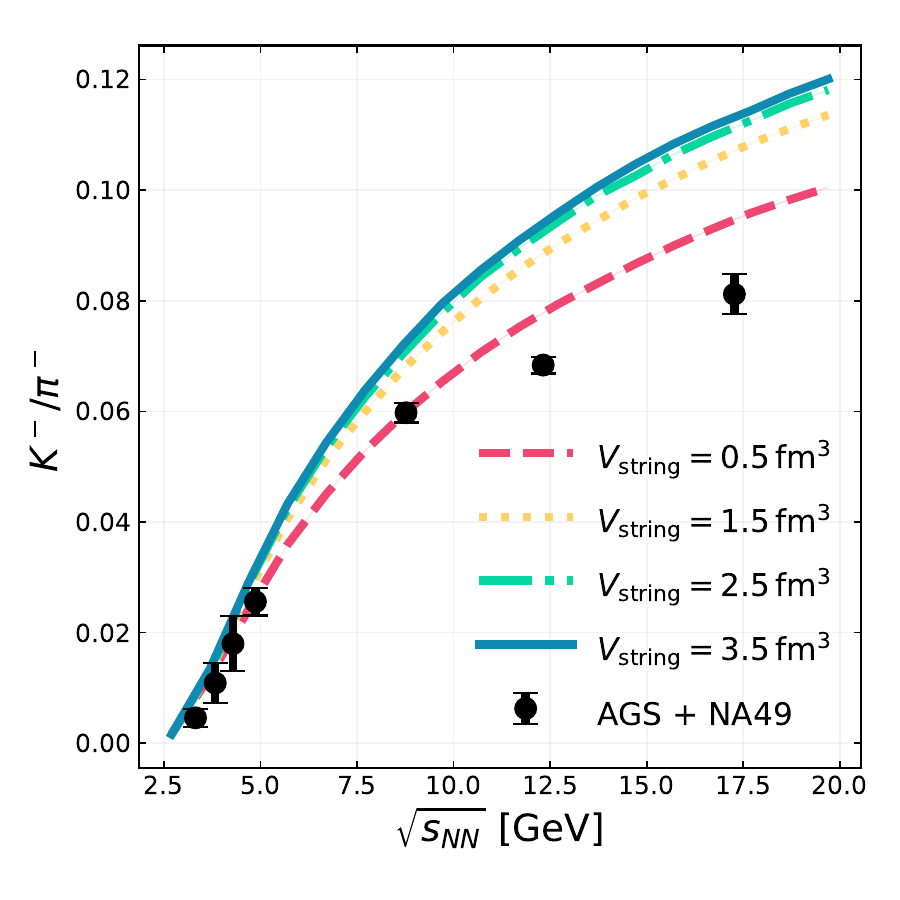}
\caption{Collision-energy dependence of the ratio of mean
$K^-$ and $\pi^-$ multiplicities in full phase space
for central Pb–Pb collisions. The different curves
correspond to different string volumes.
Experimental data from AGS and NA49 are taken from
Ref.~\cite{NA49:2002pzu}.}
\label{fig:k_pi_minus-volume-scan}
\end{figure}

Since charm production in color strings is heavily suppressed without rope formation and perturbative processes in SMASH at the energies considered here are negligible or absent (see Fig.~\ref{fig:tot_xs}), there is no charm baseline to constrain the string volume.
Therefore, to constrain charm production, we compare with the kaon-to-pion ratios shown in Figures~\ref{fig:k_pi_plus-volume-scan} and~\ref{fig:k_pi_minus-volume-scan}. Since both strangeness and charm production are enhanced by an increased string tension, strangeness observables, such as kaon-to-pion ratios, provide experimental constraints on the magnitude of the possible charm enhancement.

Before making further predictions for charm production, it is therefore necessary to constrain the string volume $V_{\mathrm{string}}$. Figures~\ref{fig:k_pi_plus-volume-scan} and~\ref{fig:k_pi_minus-volume-scan} show the collision-energy dependence of the $K^+/\pi^+$ and $K^-/\pi^-$ ratios, respectively, for central Pb–Pb collisions. The different curves correspond to string volumes ranging from $V_{\mathrm{string}}=\SI{0.5}{fm^3}$ to $\SI{3.5}{fm^3}$. As expected, larger string volumes lead to increased strangeness production. Compared to experimental data, string volumes between $\SI{0.5}{fm^3}$ and $\SI{3.5}{fm^3}$ are roughly compatible with the measurements. However, the $K^-/\pi^-$ ratios favor smaller volumes, while the $K^+/\pi^+$ ratios favor larger volumes. In particular, the $K^+/\pi^+$ ratio is well described with a string volume of $\SI{3.5}{fm^3}$, although the experimental data increasingly favors smaller string volumes with increasing collision energy.

Currently, SMASH+Ropes uses simple volume addition, which may overestimate the string fusion probability. A more detailed description of string overlap and considerations of how an increased string tension may affect the string volume could improve the description of experimental data. Further effects could include thermalization into a fluid, which would disrupt rope formation, or string repulsion through string shoving~\cite{Bierlich:2016vgw}.

In conclusion, the comparison with kaon-to-pion ratios suggests that a string volume of up to $\SI{3.5}{fm^3}$ provides a reasonable choice for estimating charm production at collision energies up to $\sqrt{s_{NN}}=\SI{10}{GeV}$, while for higher energies a smaller string volume  around $\SI{0.5}{fm^3}$ might be more realistic. In the following we will use these two values for the string volume to provide an estimate for the model uncertainty.

\section{Predictions for charm production}
Before presenting predictions for charm production, it is important to note that SMASH can produce charm through Pythia’s MPI framework, including gluon fusion ($gg \rightarrow c\bar{c}$) and quark–antiquark annihilation ($q\bar{q} \rightarrow c\bar{c}$). However, since the hard-scattering cross section is negligible at the collision energies considered here (see Fig.~\ref{fig:tot_xs}), charm production proceeds almost exclusively through string fragmentation with an increased string tension. Furthermore, charm production in Pythia, as studied using Angantyr in Ref.~\cite{Merzlaya:2024cbt}, is approximately two orders of magnitude below the experimental data and therefore represents a negligible contribution to the total charm yield.

\begin{figure}[t]
    \centering
    \includegraphics[width=\columnwidth]{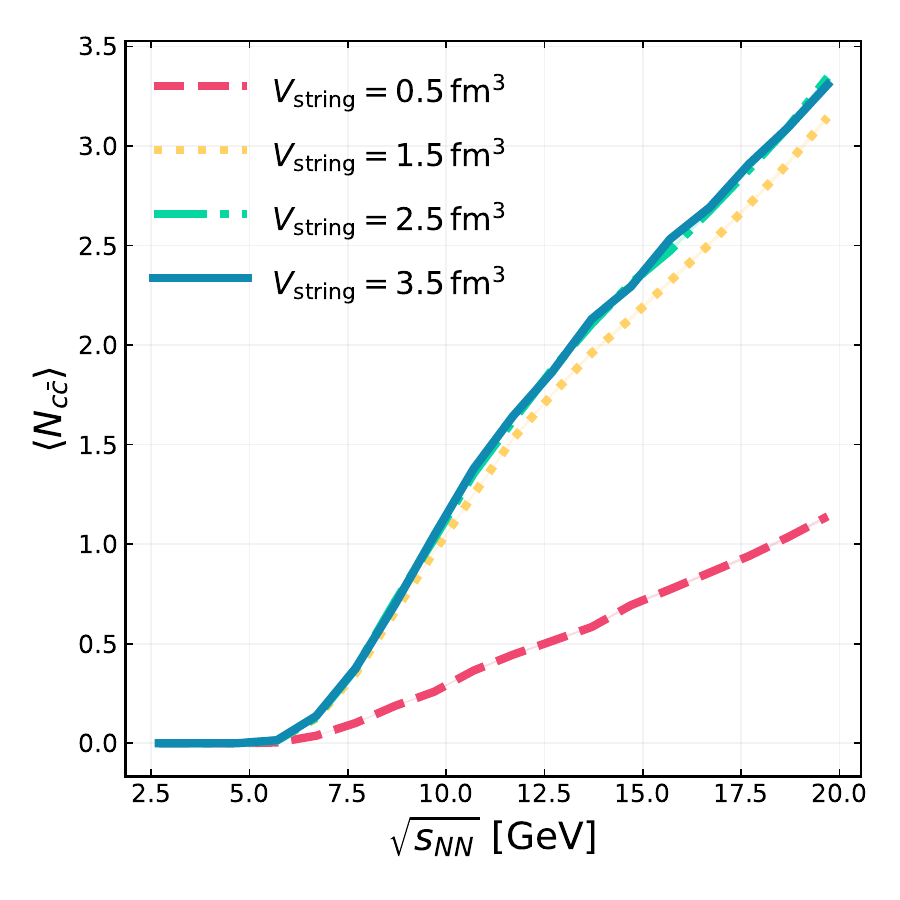}
    \caption{Collision-energy dependence of the mean $c\bar c$
    multiplicity in full phase space for central Pb--Pb
    collisions. The different curves correspond to
    different string volumes.
    }
    \label{fig:ccbar-volume-scan}
\end{figure}

Let us start with the total production of charm quarks to explore the overall magnitude of the effect obtained by incorporating string fusion into the dynamics. In Fig.~\ref{fig:ccbar-volume-scan} we show the collision-energy dependence of the mean $c\bar c$ multiplicity in full phase space for central Pb--Pb collisions. The different curves correspond to different string volumes. One observes a strong increase in the total charm production when string fusion is taken into account. One should note that with increasing string volume the effective increase saturates, because a further enhancement of the string tension leads to a saturation of the production probability.

\begin{figure}[t]
    \centering
    \includegraphics[width=\columnwidth]{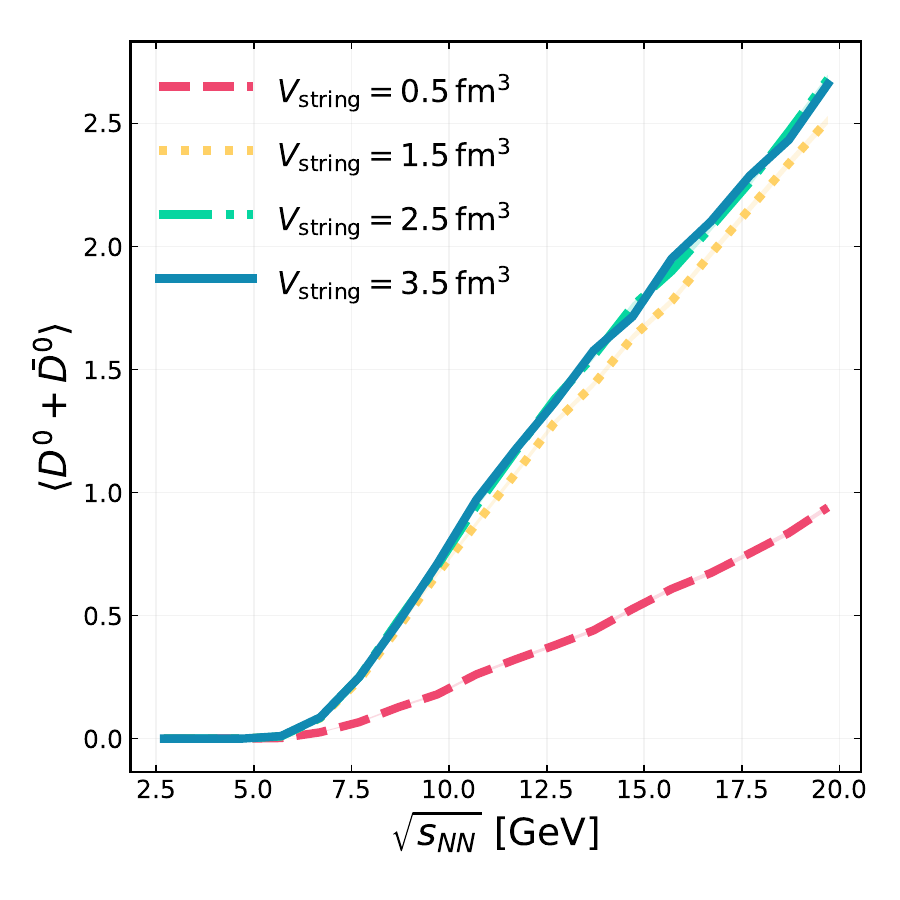}
    \caption{Collision-energy dependence of the mean $D^0+\bar{D}^0$
    multiplicity in full phase space for central Pb--Pb
    collisions. The different curves correspond to
    different string volumes.
    }
    \label{fig:d0mult-volume-scan}
\end{figure}

After the hadronization of the charm quarks, most of them will be transformed into $D$ and $\bar D$ mesons by hadronization. Here we focus on neutral $D$ mesons, as they can be experimentally well detected (e.g. by the NA61 experiment in this energy range) using their decay $D^0 \to K^-\pi^+$ (and charge conjugate). In Fig. \ref{fig:d0mult-volume-scan} we show the collision-energy dependence of the mean $D^0 + \bar{D}^0$ multiplicity in full phase space for central Pb--Pb collisions. The different curves correspond again to the  different string volumes introduced above. The initially increased charm yield is directly reflected in a strong enhancement of the D-meson production. One should note that this increase in D-meson production is mainly restricted to low/moderate transverse momenta due to the non-perturbative nature of the effect. Therefore it will not contradict the good description of charm data at high transverse momenta using FONLL calculations, but can solve the tension with the data (at the moment only available from ALICE at LHC) at low transverse momenta, where the FONLL calculations strongly underpredict the measured data.

\begin{figure}[t]
    \centering
    \includegraphics[width=\columnwidth]{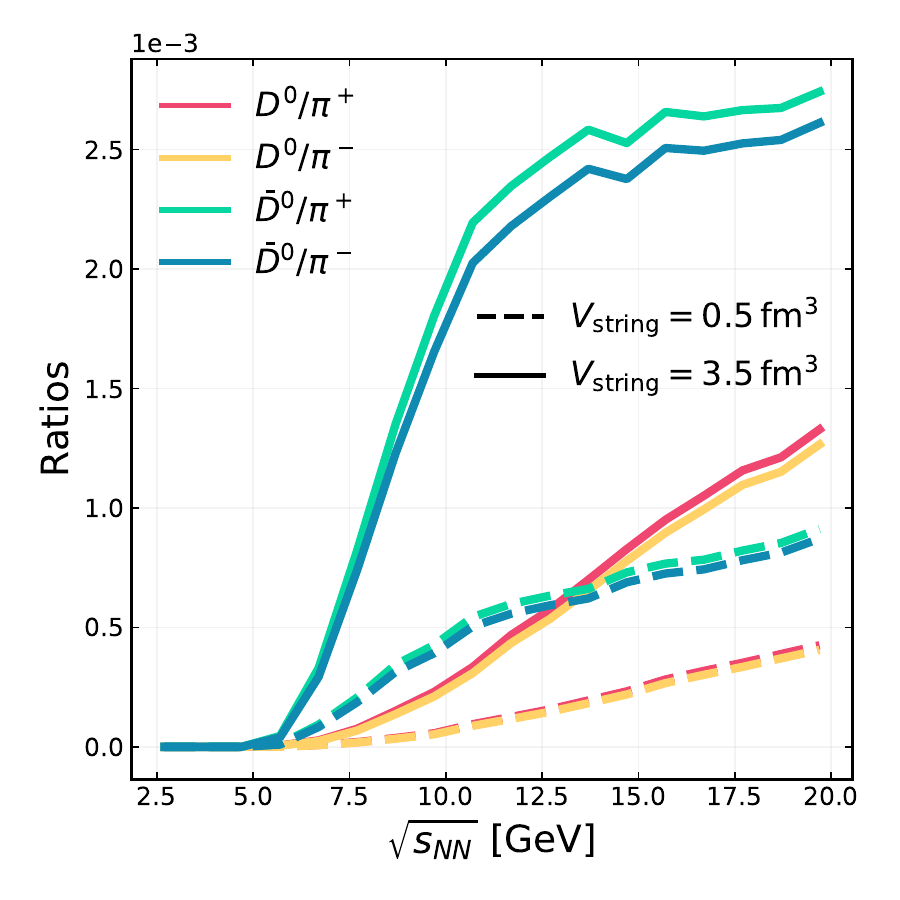}
    \caption{Collision-energy dependence of the ratios of mean
    neutral D-meson multiplicities to pion multiplicities
    in full phase space for central Pb--Pb collisions.
    The different colors correspond to different
    neutral D-meson-to-pion combinations.
    Solid and dashed lines represent string volumes of
    $V_{\mathrm{string}} = 3.5\,\mathrm{fm}^3$ and
    $V_{\mathrm{string}} = 0.5\,\mathrm{fm}^3$, respectively.
    }
    \label{fig:uncharged_dmeson_ratios}
\end{figure}

\begin{figure}[t]
    \centering
    \includegraphics[width=\columnwidth]{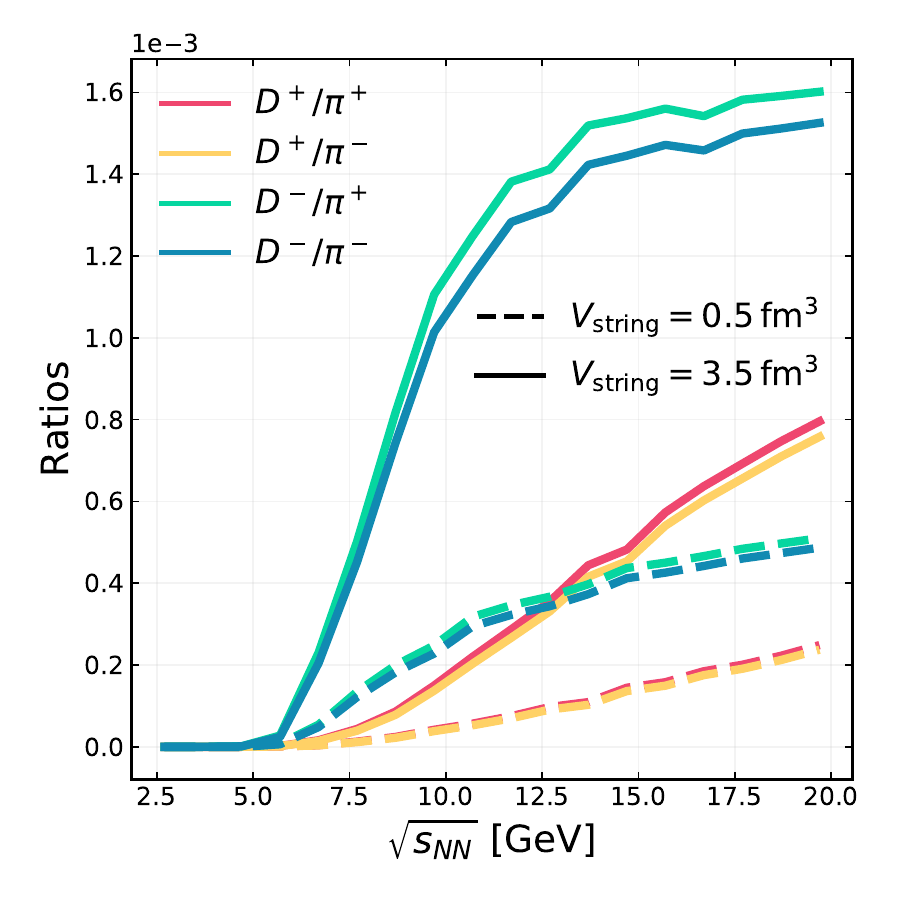}
    \caption{Collision-energy dependence of the ratios of mean
    charged D-meson multiplicities to pion multiplicities
    in full phase space for central Pb--Pb collisions.
    The different colors correspond to different
    charged D-meson-to-pion combinations.
    Solid and dashed lines represent string volumes of
    $V_{\mathrm{string}} = 3.5\,\mathrm{fm}^3$ and
    $V_{\mathrm{string}} = 0.5\,\mathrm{fm}^3$, respectively.
    }
    \label{fig:charged_dmeson_ratios}
\end{figure}

To relate charm production to light-hadron production, we normalize the different D-meson yields by the corresponding charged-pion yields. Figures~\ref{fig:uncharged_dmeson_ratios} and \ref{fig:charged_dmeson_ratios} show the collision-energy dependence of these ratios for neutral and charged D mesons, respectively, considering the two different string volumes $\SI{0.5}{fm^3}$ and $\SI{3.5}{fm^3}$ as the boundaries of the expected effect.

For both charged and neutral D mesons, the energy dependence differs significantly between the two string volumes. At $V_{\mathrm{string}}=\SI{3.5}{fm^3}$, the ratios involving anticharmed mesons exhibit a pronounced, nearly step-like increase between center-of-mass energies of approximately \SI{6}{GeV} and \SI{11}{GeV}, followed by a considerably weaker increase at higher energies. In contrast, the ratios involving charmed mesons increase more gradually over the investigated energy range. For $V_{\mathrm{string}}=\SI{0.5}{fm^3}$, all ratios exhibit a smoother energy dependence, without the pronounced enhancement observed for the larger string volume.

The strong enhancement of the anticharmed meson ratios for the larger string volume is consistent with the onset of string fusion, which increases the effective string tension and enhances charm production relative to light-hadron production. At higher collision energies, the weaker increase suggests that the relative enhancement becomes less pronounced.

Guided by the constraints on the string volume obtained from the kaon-to-pion ratios, these results suggest a potentially non-monotonic energy dependence of charm production relative to light-hadron production. In particular, a transition from the behavior associated with the larger string volume to that associated with the smaller string volume could lead to a pronounced maximum in the charm-to-pion ratios at intermediate collision energies.
Furthermore, the average yields of neutral $D$ mesons for different centrality classes are presented in Tables~\ref{tab:d0_centrality_8p8} and \ref{tab:d0_centrality_16p8}. The results exhibit a strong centrality dependence, with significantly larger yields in more central collisions.

Experimental data on charm production at the collision energies considered here remain limited. In Ref.~\cite{Merzlaya:2024cbt}, the NA61/SHINE Collaboration reported the first measurement of neutral $D$-meson production in nucleus–nucleus collisions at SPS energies. For the 20\% most central Xe+La collisions at $\sqrt{s_{NN}}=\SI{16.8}{GeV}$, the full-phase-space mean multiplicity was estimated as $\langle D^0+\bar D^0\rangle = 0.218 \pm 0.039 \pm 0.060$, $0.303 \pm 0.054 \pm 0.074$, and $0.300 \pm 0.052 \pm 0.075$, using AMPT, PHSD, and PYTHIA/Angantyr for the extrapolation, respectively. The first and second uncertainties are statistical and systematic, respectively. There is also an upper bound produced in Ref.~\cite{NA49:2005hwi} from central Pb+Pb collisions at a CM energy of $\SI{17.3}{GeV}$ which limits an average yield of 2.4 neutral $D$ mesons per event at 98\% CL. The SMASH+Ropes yields listed in Table~\ref{tab:d0_centrality_16p8} for central Pb+Pb collisions at $\sqrt{s_{NN}}=\SI{16.8}{GeV}$ are 0.681 and 2.12 for string volumes of $\SI{0.5}{fm^3}$ and $\SI{3.5}{fm^3}$, respectively.

The latter value sits just under the limit given by Ref.~\cite{NA49:2005hwi} while the yield for the lower string volume is comparable with the experimental values from the Xe+La system reported from Ref.~\cite{Merzlaya:2024cbt}. Therefore, the extension of SMASH+Ropes to include a charm enhancement due to the increased string tension in the same way as strangeness enhancement gives yields of neutral $D$-mesons that are compatible with current experimental results. Note again that contributions from pQCD (following Angantyr/PYTHIA) are negligible at these energies \cite{Merzlaya:2024cbt}.

Given that the string volume controls charm enhancement, one could in principle fit this parameter directly to the experimental charm yields. However, as discussed in Sec.~\ref{sec:charm_constraints}, experimental constraints from strangeness production already limit the range of allowed string volumes. Consequently, the predicted charm yields are not independently fitted to charm measurements but instead follow from a model constrained by strangeness production. This strengthens the predictive power of the approach and provides an independent test of the extension of SMASH+Ropes to charm production.

\begin{table}[htbp]
\centering
\caption{Average $D^0+\bar{D}^0$ multiplicity per event in Pb--Pb collisions at $\sqrt{s_{NN}}=8.8\,\mathrm{GeV}$ for different string volumes. Parentheses indicate relative statistical uncertainties in percent.}
\begin{tabular}{l
                S[table-format=1.2e-1] @{\hspace{0.4em}} r @{\hspace{1.5em}}
                S[table-format=1.2e-1] @{\hspace{0.4em}} r}
\toprule
 & \multicolumn{4}{c}{$\langle D^0+\bar{D}^0\rangle$} \\
\cmidrule(l){2-5}
Centrality & \multicolumn{2}{c}{$V_{\mathrm{string}}=0.5\,\mathrm{fm}^3$} & \multicolumn{2}{c}{$V_{\mathrm{string}}=3.5\,\mathrm{fm}^3$} \\
\midrule
0--5\%   & 1.37e-1 & $(\pm 1.2\%)$ & 5.04e-1 & $(\pm 0.6\%)$ \\
5--10\%  & 7.81e-2 & $(\pm 1.6\%)$ & 3.26e-1 & $(\pm 0.8\%)$ \\
10--20\% & 4.53e-2 & $(\pm 2.2\%)$ & 2.12e-1 & $(\pm 1.0\%)$ \\
20--30\% & 2.01e-2 & $(\pm 3.2\%)$ & 1.24e-1 & $(\pm 1.3\%)$ \\
30--40\% & 7.06e-3 & $(\pm 5.5\%)$ & 6.32e-2 & $(\pm 1.8\%)$ \\
40--50\% & 1.64e-3 & $(\pm\phantom{.}11\%)$ & 2.92e-2 & $(\pm 2.7\%)$ \\
50--60\% & 3.37e-4 & $(\pm\phantom{.}25\%)$ & 9.38e-3 & $(\pm 4.7\%)$ \\
\bottomrule
\end{tabular}
\label{tab:d0_centrality_8p8}
\end{table}

\begin{table}[htbp]
\centering
\caption{Average $D^0+\bar{D}^0$ multiplicity per event in Pb--Pb collisions at $\sqrt{s_{NN}}=16.8\,\mathrm{GeV}$ for different string volumes. Parentheses indicate relative statistical uncertainties in percent.}
\begin{tabular}{l
                S[table-format=1.2e-1] @{\hspace{0.4em}} r @{\hspace{1.5em}}
                S[table-format=1.2e-1] @{\hspace{0.4em}} r}
\toprule
 & \multicolumn{4}{c}{$\langle D^0+\bar{D}^0\rangle$} \\
\cmidrule(l){2-5}
Centrality & \multicolumn{2}{c}{$V_{\mathrm{string}}=0.5\,\mathrm{fm}^3$} & \multicolumn{2}{c}{$V_{\mathrm{string}}=3.5\,\mathrm{fm}^3$} \\
\midrule
0--5\%   & 6.81e-1 & $(\pm 0.6\%)$ & 2.12    & $(\pm 0.3\%)$ \\
5--10\%  & 4.53e-1 & $(\pm 0.7\%)$ & 1.47    & $(\pm 0.4\%)$ \\
10--20\% & 2.75e-1 & $(\pm 0.9\%)$ & 9.78e-1 & $(\pm 0.5\%)$ \\
20--30\% & 1.36e-1 & $(\pm 1.2\%)$ & 5.64e-1 & $(\pm 0.6\%)$ \\
30--40\% & 6.03e-2 & $(\pm 1.9\%)$ & 3.10e-1 & $(\pm 0.8\%)$ \\
40--50\% & 1.77e-2 & $(\pm 3.5\%)$ & 1.47e-1 & $(\pm 1.2\%)$ \\
50--60\% & 3.90e-3 & $(\pm 7.4\%)$ & 5.23e-2 & $(\pm 2.0\%)$ \\
\bottomrule
\end{tabular}
\label{tab:d0_centrality_16p8}
\end{table}

\section{Summary}
We have explored the potential influence of color rope formation on the production of charm quarks. We have shown that one may expect a substantial enhancement of charm quark production due to this additional non-perturbative process over the standard pQCD FONLL results. This enhancement is mostly restricted to the low transverse momentum regime and may resolve the tension of the ALICE data with the FONLL calculations at low $p_T$. We have presented specific predictions for the energy dependence of D-meson production in the RHIC-BES and SPS energy regime, which could be measured by the NA61 experiment in the near future. An especially interesting result is that the enhancement due to color rope formation has a rather strong energy dependence (nearly step-wise) at low energies, because string fusion increases strongly with the number of strings and the collision energy, while at higher energies the enhancement of charm production over pions saturates as a further increase of the effective string tension does not lead to an even higher probability of charm production.

Constrained by strangeness production, we predicted average neutral $D$-meson yields for several centrality classes for Pb+Pb collisions at the CM energies of $\SI{8.8}{GeV}$ and $\SI{16.8}{GeV}$ which is relevant for the NA61 experiment. Given the preference for a decreased string volume as a function of collision energy, we expect the average neutral $D$-meson yield to be $\langle D^0+\overline D^0\rangle= 0.504$ at the lower energy and for the higher energy between $\langle D^0+\overline D^0\rangle=0.681$ and $2.12$. If the effective string volume changes as a function of energy, there could even be a non-monotonic behavior in ratios of $D$-mesons to pions but also in total yields as functions of collision energy.

\begin{acknowledgments}

Financial support from the F\&E program of GSI is gratefully acknowledged. Computational resources were provided by the IT Department at the GSI Helmholtzzentrum für Schwerionenforschung, Darmstadt, Germany.
\end{acknowledgments}

\appendix
\section{Software Versions}
\label{app:software_versions}

The versions and tags of the software components used in this work are summarized in Table~\ref{tab:model-versions}. These reflect the state of the respective codes at the time the simulations were performed. However, both Pythia and SMASH were modified: SMASH to include rope formation as described in Ref.~\cite{Rosenkvist:2026wyp}, and Pythia to include an additional parameter for charm suppression.

\begin{table}[htbp]
\centering
\caption{Versions of the physics models and tools used in this work.}
\label{tab:model-versions}
\begin{tabular}{lc}
\toprule
\textbf{Physics Software} & \textbf{Version / Tag} \\
\midrule
Pythia & 8.317 \cite{bierlich2022comprehensiveguidephysicsusage} \\
SMASH & 3.4.2 \cite{weil_2026_23417573} \\
\bottomrule
\end{tabular}
\end{table}

In addition, all data were analyzed using the BRASS framework~\cite{rosenkvist_2026_23975944}, a general-purpose C++/Python toolkit originally developed for the binary output format used by SMASH.

\bibliography{ref}

\end{document}